\documentclass[twocolumn,twoside,slac]{revtex4}
\usepackage{graphicx}
\usepackage{fancyhdr}
\begin{document}

\title{NICOS System of Nightly Builds for Distributed Development}


\author{A. Undrus}
\affiliation{Brookhaven National Laboratory, Upton, NY 11973, USA}

\begin{abstract}
NICOS, NIghtly COntrol System, is a flexible tool for coordination of
software development in large-scale projects. It manages the multi-platform 
nightly builds based on the recent versions of software packages, 
tries to compensate for technical failures, tests the newly built 
software, identifies possible problems, and makes
results immediately available to developers spread over different
institutions and countries. The NICOS nightly build services ensure that 
new software submissions are consistent and provide expected results. 
The NICOS tool was developed to coordinate the efforts of more than 
100 developers from 34 countries for the ATLAS 
project at CERN 
and can be easily adapted for other large software projects. 

\end{abstract}

\maketitle

\thispagestyle{fancy}


\section{Introduction}

The software projects for High Energy Physics experiments
are international with often several hundreds programmers
from institutions world-wide involved. The coordination
of distributed software development is an important factor
of success. The automated nightly builds become a 
major component in collaborative software organization
and management. In multi-person, multi-platform environment
they provide a fast feedback to developers on new code
submissions and facilitate collective work on different
or same parts of software.

NICOS~\cite{nicos-url} was originally created 
for the ATLAS experiment~\cite{atlas-exp} and evolved in a versatile
nightly builds system. It operates on UNIX-like
platforms and works with known release management
tools, such as SCRAM~\cite{scram-url} and CMT~\cite{cmt-url}.
The NICOS nightly builds become a media
in which advanced programmers perform code development
and make this development more effective and high-quality.
Currently the NICOS tool is also used in
the LHC Computing Grid Project~\cite{lcg}.

\section{Goals and Design Principles}
The task of NICOS is to provide a nightly build system
that 
\begin{itemize}
\item works with known software release tools,
\item provides options for version management and 
number of releases in a cycle, 
\item performs testing of the newly built software, 
\item informs programmers about results with
dynamic web pages and personal e-mails. 
\end{itemize}
The design strategy is based on a goal to make
the system flexible, portable, stable, and easy to use.  

Flexibility is attained by organizing NICOS in a modular 
way so that each module is responsible for a certain
step and independently described in the NICOS configuration
file. The modular structure allows to create the nightly builds
framework that allows to plug in build, test, validation and
other external tools. For the task of software management NICOS
relies on external release tools~\cite{scram-url,cmt-url}.

NICOS is based on PERL scripts that makes possible  porting  to
UNIX and Windows systems. NICOS creates web pages that
can be viewed with major web browsers, such as Netscape,
Internet Explorer, and Opera. 

The NICOS system include the peer process that
controls the execution of modules. It detects problems
and makes an attempt to rerun failed modules. This feature
allows to compensate for temporary technical problems
and achieve maximum stability of the nightly builds.

NICOS is easy to use for both administrators and software
developers. The NICOS project configuration is stored in 
one unique text file named {\tt nicos\_cache}. 
The versions of packages for the nightly builds can be
specified in the NICOS database files or supplied by
an external tool. A fast feedback to developers is one of the
most important goals of nightly builds.
NICOS automatically posts the information
about the progress of nightly builds, identifies problems, and 
creates the web pages with build results. The authors of failed packages
can receive automatic e-mail notifications if desired. 

\section{NICOS structure}

\begin{figure*}[t]
\centering
\includegraphics[width=135mm]{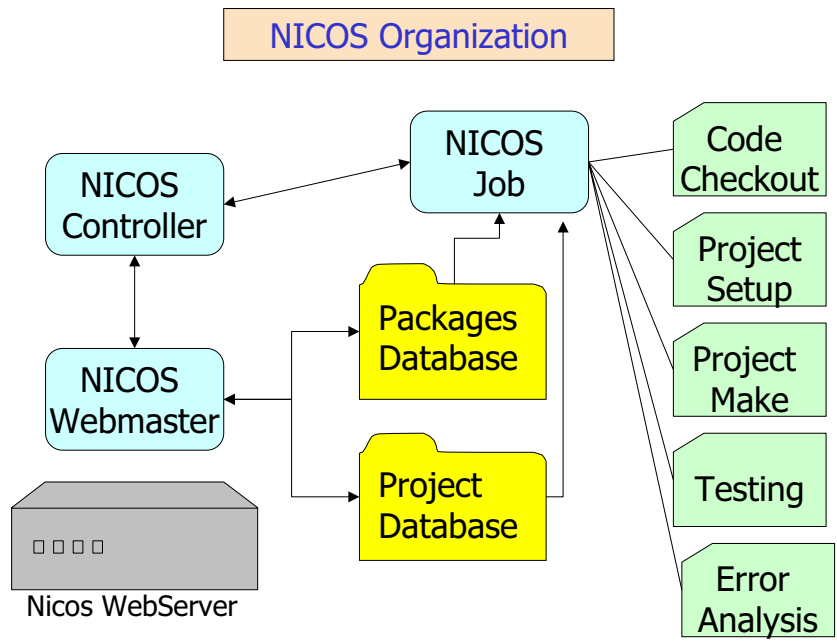}
\caption{Nicos modular organization.} \label{organiz}
\end{figure*}

NICOS organization
is shown in Figure ~\ref{organiz}. 
The main process, NICOS Controller, can be run 
at scheduled times by system tools, such as cron. It runs
the NICOS job modules and directs the status information to
the NICOS project web page.
If desired, the previous nightly release is preserved
and new release is built in a separate area. The number
of releases in a cycle is determined by the NICOS
administrator.
NICOS supports multi-platform builds (starting version 0.2)
sharing code sources. The main build process checks out
code and triggers the start of builds on other platforms.

In case of failure the NICOS
Controller tries to make a restart from the point of failure.
If the NICOS job runs overtime (typically more than 24 hours)
it is automatically destroyed.

There are eight major modules in the NICOS job. They are
able to serve as the interfaces to external tools, such as CVS,
test, and release management tools.

\begin{itemize}
\item NICOS configuration. At this step the project parameters
are read from the {\tt nicos\_cache} file.

\item Release tool setup. The release tool (such as SCRAM, CMT)
is setup with commands specified in {\tt nicos\_cache}.

\item Code checkout from the CVS repository specified in {\tt nicos\_cache}.
The packages tags are retrieved from the NICOS database file or
supplied by an external tool.

\item Project setup. At this step the project setup
commands can be specified.
\item Project build. The release tool should insert separators
between the packages builds outputs with the packages names.
The unique string from the separators should be specified
for this step. After the build NICOS cuts out and writes the logfiles
with the build outputs for individual packages to the {\tt Log} area
of the project.

\item Unit tests and Integrated tests. The test tools,
such as OVAL~\cite{oval} or CppUnit~\cite{cpp}
are plugged in, as desired.
The outputs are forwarded
to the {\tt TestLog} area of the project.

\item Error Analysis. In the build output NICOS searches
for critical patterns indicating {\tt make} problems,
such as {\tt 'No rule to make target'} or
{\tt 'Symbol referencing errors'}. For tests,
success is determined by the value returned by
test tools.

\item Creation of the web pages with the summary of build results
and e-mail notification about problems. 

\end{itemize}

\begin{figure*}[t]
\centering
\includegraphics[width=135mm]{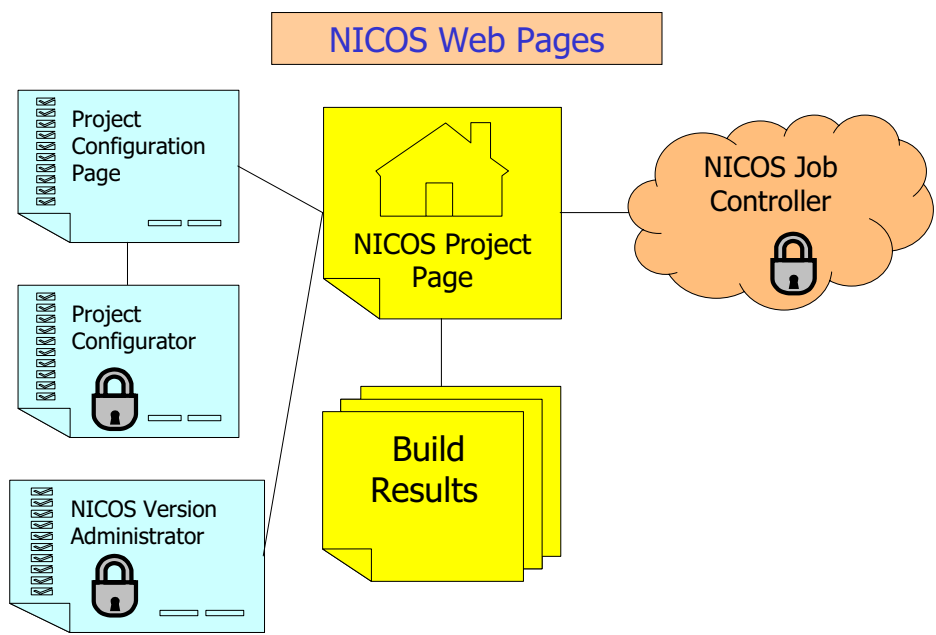}
\caption{Nicos web pages. Secured pages are marked with lock sign.}
\label{web}
\end{figure*}

The progress and results of NICOS builds are reflected
on the dynamically created web pages. Its structure is shown
in Figure ~\ref{web}. The NICOS project web page shows the 
major parameters, such as project area, compiler options,
and list of releases available for supported platforms.
For building releases, the currently performed step
is displayed. The NICOS project web page contains links
to the pages with detailed results for individual releases
and pages showing the NICOS database and configuration
parameters.

Optionally the collection of PHP scripts, ``NICOS Webmaster'',
 can be used for handling the administration of NICOS over the WWW
(available in 0.2 version of NICOS). This tool
can modify the project configuration ({\tt ``Project
Configurator''}) and  NICOS database of package versions
({\tt ``Version Administrator''}). In addition it
is possible to schedule or stop
builds from the {\tt Job Controller} page.

\section{Configuration Management}

 The configuration parameters
for the NICOS job is collected in the XML-like {\tt nicos\_cache} files.
The steps of building process is associated with its markup 
tag. Every tag consist of a tag name, sometimes followed by
an optional list of tag attributes that are placed between
brackets ( $<$ and $>$ ). The markup tag is followed by the 
commands, including environment definitions, needed to be
executed at the step. The example of a markup tag below
is for the step of build. The attribute defines the
directory name (relative to the project area)
where the build should be performed. The tag is 
followed by command {\tt scram b} that builds a 
SCRAM based project.

\begin{verbatim}
     <project build dir=src>
     scram b
\end{verbatim}

 NICOS checks out versions of packages as specified 
in the NICOS database file. 
Each line of this text file contains the name of package, 
tag of package, and e-mail addresses of the developers
separated by spaces. The exact tag of package or
option for tag calculation can be indicated.
The options include the selection of recent submission,
recent tagged submission, recent version in the ``official''
format: {\tt <package name>-[0-9][0-9]-[0-9][0-9]-[0-9][0-9]}.
The NICOS database file can be indicated in {\tt nicos\_cache}
or dynamically created by a plug-in script. The latter
option is useful for large collaborations where
special version management tools are used~\cite{tgcol}.

\section{Status and Plans}

The 0.1 version of NICOS is available since March 2003
and provides basic functionalities. 
New 0.2 version (scheduled for September 2003) will
contain important improvements (e.g. 
multi-platform support) and additions such as
the PHP script collection for administration over
the WWW. The long term plans center around improving
simplicity of use and portability (including
the support of NICOS for Windows OS).

\section{Conclusions}
NICOS (NIghtly COntrol System) is currently 
successfully used by several software
projects at CERN (e.g. ATLAS, LCG). The NICOS features
such as the instantaneous error analysis, dynamic
informative web pages, and e-mail notifications about problems
are proved to be useful for collaborative software organization
and management. 
As experience gains, NICOS undergoes continuous 
improvements and evolutions.
 
\begin{acknowledgments}
The author wishes to thank members of the ATLAS Software Infrastructure 
team, the LCG SPI group and Physics Applications 
group at BNL for many valuable advices and
useful discussions.
\end{acknowledgments}


\begin{thebibliography}{99}   % Use for  1-9  references
%\begin{thebibliography}{99} % Use for 10-99 references

\bibitem{nicos-url}
The NICOS home page is http:
//www.usatlas.bnl.gov/computing/software/nicos

\bibitem{atlas-exp}
http://atlas.web.cern.ch/Atlas

\bibitem{scram-url} 
http://cmsdoc.cern.ch/Releases/
SCRAM/V0\_19\_8/doc/html

\bibitem{cmt-url}
http://www.cmtsite.org

\bibitem{lcg}
http://lcgapp.cern.ch/project

\bibitem{oval}
http://polywww.in2p3.fr/cms/software/oval

\bibitem{cpp}
http://sourceforge.net/projects/cppunit

\bibitem{tgcol}
See for instance description of Tag Collector tool in  
S. Albrand, J. Collot, J. Fulachier, ``The AMI Database Project'', CHEP2003 

\end{thebibliography}
\end{document}